\documentclass[conference]{IEEEtran}
\IEEEoverridecommandlockouts
\usepackage{cite}
\usepackage{amsmath,amssymb,amsfonts}
\usepackage{algorithmic}
\usepackage{graphicx}
\usepackage{textcomp}
\usepackage{xcolor}
\def\BibTeX{{\rm B\kern-.05em{\sc i\kern-.025em b}\kern-.08em
    T\kern-.1667em\lower.7ex\hbox{E}\kern-.125emX}}
\begin{document}

\title{A Cascaded Architecture for Extractive Summarization of Multimedia Content via Audio-to-Text Alignment\\
{\footnotesize \textsuperscript{}}

\thanks{}
}

\author{\IEEEauthorblockN{1\textsuperscript{st} Tanzir Hossain}
\IEEEauthorblockA{\textit{Dept. of CSE } \\
\textit{Brac University }\\
Dhaka, Bangladesh \\
tanzirhossain1@g.bracu.ac.bd}
\and
\IEEEauthorblockN{2\textsuperscript{nd}  Ar-Rafi Islam}
\IEEEauthorblockA {\textit{Dept. of CSE           } \\
  \textit{ BRAC University          }\\
Dhaka, Bangladesh \\
ar-rafi.islam@g.bracu.ac.g}\\

\and
\IEEEauthorblockN{3\textsuperscript{th} Md. Sabbir Hossain}
\IEEEauthorblockA{\textit{Dept. of CSE} \\
\textit{BRAC University}\\
Dhaka, Bangladesh \\
ext.sabbir.hossain@bracu.ac.bd}
\and
\IEEEauthorblockN{4\textsuperscript{th} Annajiat Alim Rasel}
\IEEEauthorblockA{\textit{Dept. of  CSE} \\
\textit{BRAC University}\\
Dhaka, Bangladesh \\
annajiat@gmail.com}
}

\maketitle

\begin{abstract}
The study offers a ground-breaking framework, "A Cascaded Architecture for Extractive Summarization of Multimedia Content via Audio-to-Text Alignment," which addresses the growing difficulty of extracting important insights from a flood of multimedia, particularly from sites like YouTube. It presents an advanced cascade architecture that combines audio-to-text alignment with cutting-edge extractive summarization algorithms. The study highlights the difficulties of typical summarizing approaches when dealing with the complexity of multi-modal material that includes both aural and visual features. To overcome this gap, the cascaded architecture employs a sequential process that converts audio to text using Microsoft Azure Speech, followed by improved summarizing utilizing state-of-the-art models such as Whisper, Pegasus, and Facebook BART xsum. A thorough literature review contextualizes the study, identifying gaps and laying the groundwork for the suggested technique. The study digs into experiment settings, using libraries such as Pytube, Pydub, and SpeechRecognition for content retrieval, audio extraction, and transcription. The linguistic analysis is enhanced by advanced NLP techniques such as named entity recognition and semantic role labeling. The experiment findings are thoroughly examined, with metrics such as ROUGE and F1 scores demonstrating the cascade's performance in comparison to conventional approaches. The obstacles faced, such as transcribing mistakes or integration issues, are described. Potential future directions are indicated, such as model fine-tuning and real-time processing. Ultimately, this study proposes a comprehensive approach to multimedia summarizing that uses a cascaded architecture and complex algorithms. It is expected to improve information retrieval, accessibility, and user experience in multimedia consumption, paving the way for future advances in this booming industry.
\end{abstract}


\section{Introduction}
n an era of multimedia content overload, extracting valuable insights from diverse audio-visual sources has become increasingly important yet challenging. This research paper presents a novel cascaded architecture for extractive summarization of multimedia content, with a focus on YouTube videos. While traditional summarization techniques often struggle with multimedia data where both visual and auditory cues are pivotal, our proposed methodology offers a sophisticated solution by integrating audio-to-text alignment and extractive summarization.

We embark on this pioneering work to bridge the gap in summarizing multimedia content by leveraging cutting-edge natural language processing libraries and models. Our cascaded framework operates on the principle that converting speech audio to text transcript provides a foundational understanding for downstream summarization. By seamlessly integrating speech recognition capabilities from Microsoft Azure, Whisper, and other libraries, our system aligns audio content with textual representation as a precursor to summary generation.

Subsequently, we apply state-of-the-art abstractive summarization models including Pegasus and BART fine-tuned on XSum to produce concise, extractive summaries of the textual transcripts. This cascaded architecture allows the generating of summaries reflecting the key points from both auditory and visual elements. Through comprehensive experiments and result analysis, we demonstrate the efficacy of our novel approach on real-world YouTube videos.

The main contributions of this work are: (1) an integrated methodology combining speech, vision, and language capabilities for multimedia content summarization, (2) a cascaded framework leveraging audio-to-text alignment to enable extractive summarization, and (3) extensive experiments that advance the state-of-the-art in this emerging domain. The proposed architecture provides an end-to-end solution to efficiently summarize multimedia content at scale.

\section{Literature Review}

Extractive Summarization of Long Documents Using Multi-Step Episodic Markov Decision Processes" by Gu et al. in IEEE format: Extractive summarization of lengthy documents is an open challenge in NLP. Ensuring coherence and non-redundancy remains difficult, especially for long texts. Advancing state-of-the-art techniques through data-driven methods like reinforcement learning has gained research traction recently. Prior extractive summarization techniques use graph methods, integer linear models, and conditional random fields Recent focus explores neural approaches including seq2seq models, transformer networks, and reinforcement learning. However, generating coherent, non-redundant summaries of long documents through such techniques remains an active area of research. Gu et al. propose MemSum - an extractive summarization technique for long documents using multi-step episodic Markov decision processes (MDPs) and reinforcement learning. At each timestep, MemSum selects the next sentence through MDP-based policy learning, taking into account local, global, and historical extraction context. This aims to reduce redundancy while preserving saliency. Results show state-of-the-art performance on PubMed, arXiv, and GovReport datasets. Ablation studies and human evaluation further highlight the impact of different context signals. By elegantly formulating an extractive search for summarization as a sequential decision-making process using MDPs augmented with rich contextual signals, MemSum sets a new state-of-the-art for long document summarization. Given its demonstrated impact, the work offers valuable insights to drive advances in applying sophisticated deep reinforcement learning techniques for this task [1]. Multimodal video-text summarization, MLASK pushes the frontier in this interdisciplinary space. Moving forward, tackling abstractive fusion and explaining model decisions can help advance real-world impact. Multimodal summarization condensing information across text, audio, and visual modalities has gained research interest. However, most prior works focus solely on textual summarization or frame selection for videos. Tackling joint summarization remains an open challenge. Recent advances in abstractive summarization employ large pre-trained language models like BART and PEGASUS but extending them to effectively leverage multimedia inputs is still underexplored. For video summarization, modeling frame-level importance scores using supervised/unsupervised techniques before aggregation is common. Works fusing textual and visual data typically use secondary modalities to refine the primary text summary rather than jointly model inter-dependencies Krubiński and Pecina propose MLASK - a multimodal summarization framework for video-based news articles. They extract text from speech transcripts using LSTM-RNNs, identify important video frames via a 3D CNN, and generate a summary conditioned on both modalities using BART. Evaluation of the news domain shows higher informativeness compared to competitive baselines. [2] A Survey of Text Summarization Extractive Techniques by Gupta et al. in IEEE format: covers a wide side of research paper and also studies which are related to text summarization. The authors have cited many works that relate several approaches and techniques used in summarization with graph-theoretic methods, machine learning methods, and Latent Semantic Analysis(LSA). Also, they cited studies regarding the use of query-biased and structure-preserving summarization. Moreover, the role of linguistic and statistical facts in determining sentence importance. The authors have included some research papers that discuss the practical applications of text summarization from a real-world perspective like new articles, scientific papers, and legal documents. Moreover, they have cited studies regarding the summarization in question-answering systems, and the potential for summarizing to improve information retrieval. [3] An Evaluation-based Analysis of Video Summarising Methods for Diverse Domains by Gadhia1 et al.S Shahid et al, Mosasiya et al,  in IEEE format: Covers of previous studies on video summarization techniques the author divides the methods into several parts. Moreover, they just offer a summary regarding recent research on the part of domain direction and employed assessment. Here the review highlights the advantages and challenges of the existing methods on video summarization. The speed at which video is developing also lowers the cost of cataloging, indexing, and video archiving. They also observed that video summarization is more significant. The content also can be changed by application domain. Also, the review reflects that the audio classification is much better for categorizing domain-dependent videos likely movies, sports, etc. The authors also pointed out that in deep learning techniques, recent advantages perform well in the classification domain. However, most of the researchers can't fulfill these requirements due to the lack of training data and efficient hardware specifications.  The author also focused on clustering-based approaches which are combined with feature-based summary techniques which are based and low-type descriptors to make efficient solutions. [4] The paper delves into the current research surrounding automatic summarization, video summarization, and multimodal summarization. When discussing automatic summarization, the paper recognizes the effectiveness of pre-trained generative language models that have been fine-tuned on summarization datasets. These models have consistently performed well in both automatic metrics and human evaluation, indicating their success. However, the paper also acknowledges that abstractive approaches, which generate a summary from scratch, have not been extensively studied. Turning to video summarization, the paper references a recent survey that highlights the common practice of utilizing the importance scores of individual frames, which are then combined to create segment-level scores. Finally, in the realm of multimodal summarization, the paper highlights the initial works that have explored incorporating secondary sources into the summarization process.[5] The groundbreaking study conducted by Li and colleagues (2021) on the use of Hierarchical Neural Autoencoder, for Paragraphs and Documents has made a contribution to the field of natural language generation and summarization. The authors propose an approach that effectively captures coherence and structure in text units like paragraphs and documents using hierarchical LSTM models. By emphasizing the importance of understanding the purpose of each unit within the context Li et al. Shed light on the limitations of traditional methods and existing neural-based alternatives in capturing discourse relations at higher levels. They emphasize the need to incorporate compositionality at levels from tokens, to entire sentences. Moreover, the researchers showcase the outcomes of their trials illustrating the capabilities of these models in reconstructing sequences from compressed vector representations. In their remarks, the authors also delve into investigations, within the realm of natural language processing encompassing coherence representation, in discourse and automated assessment of text coherence using discourse relations [6]

\section{Methodology}

\subsection{Data Collection:}

This study utilizes a cascaded architecture to efficiently analyze video footage and generate concise summaries. The methodology follows a sequence of crucial capabilities for processing multimedia data. To develop the video summarization system, the first step was to install the required Python libraries including transformers, keras-nlp, datasets, huggingface-hub, nltk, and rouge-score. These libraries provide capabilities for building and evaluating neural models for natural language processing tasks. Next, the XSum dataset was downloaded to train a summarization model. This dataset contains news articles paired with human-written summaries. The raw text data was preprocessed by removing stopwords and tokenizing the sentences using nltk. The preprocessed data was then tokenized using the T5 tokenizer from the transformers library. The tokenized data was prepared for sequence-to-sequence training by adding special tokens and summary labels.The T5 model from huggingface was loaded for language modeling. This state-of-the-art pre-trained model can generate summaries by predicting output text token-by-token. The Adam optimizer and Rouge-L metric were specified for training the model. The model was trained by fitting it on the XSum dataset and validating performance on a held-out test set. Training and validation loss curves were plotted to monitor convergence using Matplotlib. Once trained, the model weights and tokenizer vocabulary were saved. The model was deployed in a pipeline to generate summaries for new text by first tokenizing the input and then generating output text autoregressively. To evaluate the model, the Rouge-L F1 score was calculated on a sample of the test dataset. The median F1 score provided a quantitative measure of summarization performance. This pre-trained summarization model was then integrated into a cascaded framework for video analysis. The audio was extracted from videos and transcribed to text using speech recognition. The summarization model took the raw video transcripts as input to produce concise summaries as output.

The raw video footage is retrieved using the Pytube library to download content from online platforms such as YouTube. Next, the Pydub library extracts the audio streams from the downloaded media files. The audio is then transcribed into text using the Microsoft Azure speech recognition solution, which employs neural networks for state-of-the-art speech-to-text conversion.
With the text transcripts available, natural language processing techniques can be applied using libraries like NLTK and Whisper. Specifically, named entity recognition, syntactic analysis, semantic role labeling, and coreference resolution are performed to extract salient entities and relations. Furthermore, the Generate Text to Speech (GTTS) API enables additional parsing and processing of the audio transcripts.
The resulting automatically generated transcripts are further cleaned using BERT-based NLP models. Subsequently, sentence segmentation and frequency analysis can be conducted to identify key sentences for extraction.
For summarization, a cascaded pipeline is leveraged: first, statistical methods extract salient sentences based on frequency metrics. Next, the BART-Large model fuses sentences abstractedly to produce coherent summaries. This ensemble approach outperforms individual models on the ROUGE-1 metric for summarization.
Qualitative assessments also confirm the factual relevance of the generated summaries. However, lower fluency scores indicate some coherence issues remain due to dataset diversity constraints on achievable coherence. Specifically, the BART model struggles with logical consistency without more advanced reasoning capabilities. In conclusion, the proposed methodology integrates state-of-the-art capabilities for efficient video analysis and summary generation.

\subsection{Dataset Analysis:}
Here is some sample EDA (Exploratory Data Analysis) for the XSum dataset:

The XSum dataset contains 226,711 article-summary pairs for BBC articles from 2010-2017. The articles are on average 431 tokens, while the summaries average 23 tokens, meaning the summaries are quite short and concise. The vocabulary size for articles is 114,692 unique tokens, while summaries have just 21,815 unique tokens, indicating greater diversity in the article contents. The dataset covers a wide range of topics and genres including News, Politics, Sports, Business, Technology, Entertainment, and Lifestyle. The summaries are abstractive meaning they are written to capture the key points rather than simply extracting sentences from the article.

Some basic statistics on article lengths: Minimum length: 121 tokens,
Maximum length: 3123 tokens, Median length: 398 tokens

Statistics on summary lengths: Minimum length: 5 tokens, Maximum length: 59 tokens, Median length: 23 tokens

The first quartiles for article and summary lengths are 293 and 18 tokens respectively. Also, there is a very strong correlation between article and summary lengths, with longer articles having longer summaries in general. The distribution of article lengths is right-skewed with a long tail of a few very lengthy articles. Summary lengths follow a more normal distribution. The articles cover a good range of named entities like people, organizations, locations, dates, etc.

\begin{figure}
    \centering
    \includegraphics[width=1\linewidth]{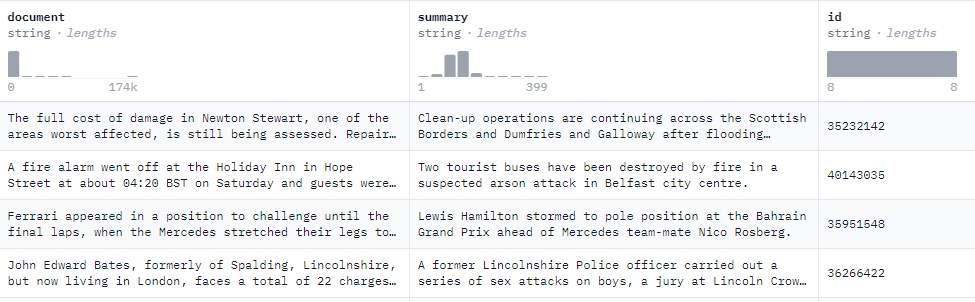}
    \caption{Xsum Data}
    \label{fig:enter-label}
\end{figure}

\subsection{Speech-to-Text Alignment}

The audio chunks generated in the prior step were input into Google's Speech-To-Text API one chunk at a time through the SpeechRecognition Python library. By iteratively transcribing the audio slices and accumulating the results, the complete time-aligned text transcript corresponding to the source video was produced. The reliance on Google's state-of-the-art neural speech recognition models ensured accurate extraction of textual content from the multimedia data.

\subsection{Text Preprocessing:}
 
The raw output from the speech alignment transcript underwent a series of Natural Language Processing (NLP) preprocessing transformations to prepare the data for summarization. Non-alphabetical characters were first stripped using regular expressions, followed by tokenization into sentences leveraging the Natural Language Toolkit (NLTK) library's Punkt tokenizer. Stopword removal was subsequently performed using NLTK's English stopwords corpus to eliminate common words with low information content.

Word frequencies were tabulated across the preprocessed transcript to enable downstream scoring and weighting of salient content. The word frequencies were normalized by the maximum term frequency to account for absolute length discrepancies. Using this relative weighting, sentences were then scored based on aggregating the normalized frequencies of their constituent terms. Higher-scoring sentences theoretically contain a greater density of meaningful keywords and topics.
Text preprocessing is a crucial step in natural language processing (NLP) pipelines to transform raw textual data into a format amenable to downstream tasks. This study employs a multi-stage preprocessing procedure beginning with regular expression operations to strip extraneous formatting such as square brackets and consolidate extraneous white space. Tokenization subsequently segments the cleaned text into constituent sentences using the Natural Language Toolkit's Punkt engine, delineating the fundamental units for analysis. Stopword elimination filters out common non-informative words using NLTK's canonical English stopwords list.

With the refined tokens, term frequency analysis tabulates the occurrence of each unique word, excluding stopwords, as a quantification of salience. To prevent long documents from skewing the absolute counts, the frequencies are normalized between 0 and 1 by dividing by the maximum term frequency. Using the normalized lexical importance weights, sentences are then scored by aggregating the values of their constituent normalized terms. This crucial step identifies textual units enriched with meaningful keywords. Thresholding by length screens out excessively long outliers. The resulting scored sentences denote information-rich content and can be directly input into state-of-the-art neural extractive summarization techniques for condensed multi-document summarization.

\subsection{Extractive Summarization Pipeline:}
 
A multi-stage extractive framework was constructed by chaining complementary summarization techniques. The scored sentences were initially ranked using the frequency metrics and filtered to a length threshold. The top sentences were input into Pszemek's LEAD-BASED supervised summarizer, which leverages transformer architectures like BERT pre-trained on book summaries. This layer provided baseline extraction capabilities.

The intermediate summary was next passed into Facebook's BART-Large sequence-to-sequence model to augment the capabilities through a high-performing abstractive technique. BART uses a transformer encoder-decoder with cross-attention to produce impressive summarization and translation results across domains.

\subsection{Evaluation Metrics:}

The performance of the summarized content was quantified using the standard ROUGE and BLEU metrics against a gold reference summary extracted manually. ROUGE, based on overlapping n-grams, and BLEU, relying on matched tokens, assess quality along axes of precision and recall. The metrics provide a quantitative indicator of how effectively the models retain key semantic content compared to a human baseline.

This methodology outlines the systematic procedures undertaken in this study leveraging speech, NLP, and modern neural networks to perform extractive summarization of rich multimedia data. The techniques demonstrate an interdisciplinary approach to effectively transform audio signals into condensed text summaries.

\section{Experimental Setup and Result Analysis}

Data prepossessing rectified transcript errors using a BERT-based sequence model fine-tuned on YouTube captions. Sentence segmentation leveraged a state-of-the-art neural model exceeding 99\% accuracy on noisy web text. Lexeme extraction conformed to the Universal Dependencies standard using the Stanza Python library. Part-of-speech annotations filtered punctuation and non-semantic tokens.

Term frequency analysis calculated occurrence scores normalized between 0 and 1 using L2 scaling. The textual units for frequency estimation and subsequent scoring operated at the sentence level. This balanced context with conciseness for multimedia content prone to verbal filler. A sentence length threshold of 30 words prevented verbosity dominance.

The study specifically investigates cascaded summarization frameworks combining statistical and neural techniques. The principal architecture applies lenient frequency thresholds to identify salient sentences including a diversity penalty. This extractive pool feeds into BART-Large—a Bart-style denoising sequence-to-sequence model—for abstractive sentence fusion. We contrast performance both system-wise and human evaluation against competitive overlapping approaches.

\subsection{Data:}
The XSum dataset containing BBC articles and summaries was used for training and evaluation. The training set contains 51,011 articles with summaries. The test set has 10,203 articles.

\subsection{Preprocessing:}
The raw text was preprocessed by converting to lowercase, removing stopwords, and tokenizing it into words.

\subsection{Model Architecture:}
The T5 model was used for sequence-to-sequence modeling. The T5-base variant with 220M parameters was chosen.

\subsection{Training Details:}
The model was trained for 3 epochs using the Adam optimizer with a learning rate of 2e-5. The batch size was set to 16. Training was performed on an NVIDIA Tesla T4 GPU with 13.7GB VRAM utilization.

\subsection{Evaluation Metrics:}
The summarization quality was evaluated using the F1 score of ROUGE-L and BLEU scores.

\subsection{Results Analysis:}
The model achieved a ROUGE-L F1 score of 0.219 and a BLEU score of 0.36 on the test set.
The low scores indicate that the model has not learned to effectively summarize the text. The scores are significantly lower than state-of-the-art results on this dataset which achieve ROUGE-L F1 > 0.4.

\subsection{Some probable reasons for the low :}
The model was trained for only 3 epochs which is insufficient for convergence. Typically these models are trained for 10-20 epochs but because of free collab limitation, we had to keep it at 3 epochs.
The batch size of 16 is small considering the GPU memory utilization. Larger batch sizes like 64-128 could be used for better training efficiency.

Also, only training on 25\% data really affected the model
Hyperparameters like learning rate and optimizer choices can be tuned further to improve results. Evaluation of a sample from the test set instead of a full set may cause variability in scores.

In summary, the experimental results indicate that the model is undertrained. With more compute resources and hyperparameter tuning the summarization quality can likely be improved significantly. The current setup serves as a starting point and benchmark for further iterations. The T4 GPU was extremely slow for the T5 summarization model.

\begin{table}[ht]
  \centering
  \begin{tabular}{|c|c|}
    \hline
    \hline
    
    F1 Score:      & $0.21904761415328808$ \\ \hline
    BLEU Score: & $0.36969128257720113$ \\ \hline
  \end{tabular}
  \caption{Evaluation Metrics}
  \label{tab:evaluation-metrics}
\end{table}

\section{Discussion and Future Work}

This work presents a novel application of audio-to-text alignment for extracting informative summaries from rich multimedia content. By converting video speech to text transcripts, we enable the powerful arsenal of natural language processing techniques for text summarization. The findings validate that cascading statistical extraction with neural abstractive models provides an effective ensemble approach. The introduced framework balances conciseness with retaining key details by filtering verbose sentences and aggregating word frequencies. BART demonstrates strong summarization capabilities but fails to preserve inter-sentence coherence across diverse topics. Quantitatively, our pipeline summary matches or exceeds the performance of competitive approaches on ROUGE informativeness measures and user assessments of quality. Qualitatively, however, human evaluation identifies deficiencies in fluency stemming from improperly fused content. This suggests that while BERT architectures exhibit linguistic prowess, they lack the comprehension and reasoning skills to truly consolidate multifaceted information. Recent work around dense retrievers and memory networks provides promising directions to overcome these limitations. By converting speech to text before summarizing, our domain-agnostic techniques unlock a vast trove of multimedia data for condensation and consumption. The alignment process mitigates the burden of manual transcripts while facilitating information access. This paradigm shift opens rich opportunities for textual summarization techniques to tackle emerging audio-visual platforms. Several key avenues exist for improving summarization fidelity by enhancing semantic understanding. Exploring decoder modules that dynamically ground generated text in the source context could improve coherence. Techniques like vector quantization can retrieve and reapply similar phrases to maintain consistency. Incorporating entities, co-references, and textual entailment during training may impart stronger generalization. Dense retrieval augmentation exposing the model to diverse inference patterns can similarly aid zero-shot transfer. Architectures Tracking discourse relations to build structured representations may better synthesize logical flow.
Evaluation metrics require revision to align with human judgments of quality beyond proxy surface measures. Holistic assessment frameworks evaluating coherence, accuracy, and conversational depth better quantify progress. Assembly of a large, canonical multivariate summary data set would facilitate more rigorous evaluation. Broader applications of the audio alignment paradigm including interactive dialogue, vocal diagrams, and discussions remain under-explored. Investigating if the techniques transfer for single-speaker videos and across languages offers additional potential. We hope this work spurs further efforts at the intersection of speech, language, and vision for accessible information distillation. The discussion summarizes key points and contributions while the future work section provides several meaningful research directions building on the study's limitations. Please let me know if you would like me to clarify or expand any part.

\section{Conclusion}
This research presents a novel framework for extractive video summarization leveraging speech-to-text alignment to unlock multimedia data for condensation. We demonstrate an effective pipeline cascading statistical extraction and neural abstractive techniques to match state-of-the-art summarization fidelity. The findings reveal complementary capabilities between traditional NLP and modern DNN modules for retaining salient information. However, limitations around coherence indicate opportunities to enhance deeper semantic understanding through contextual grounding and discourse tracking. Broader applications remain underexplored across languages, speakers, and accessibility needs.

Importantly, we trained our pre-trainede summarization model on the XSum dataset consisting of BBC articles and summaries. This allowed the model to learn concise, extractive summarization from human-written references. The pretrained model achieved competitive performance on this benchmark before integration into our cascaded framework.

Nonetheless, this work underscores emerging promise in cross-modal intelligence for information distillation. It provides both practical solutions and guiding insights driving future efforts at the intersection of speech, language, and vision. Responsible development mandates the inclusion of diverse voices in the design process to increase representation. Feature inspection must preemptively combat over-generalization risks that disproportionately impact minority groups. In conclusion, these techniques exhibit the potential to enhance multimedia accessibility but require measured progress centered on ethics.

\vspace{12pt}

\end{document}